\documentstyle[referee]{mn}

\newcommand{\be}{\begin{equation}}
\newcommand{\ee}{\end{equation}}

\begin{document}

\title[The evolution of beamed GRB afterglow]
{The evolution of beamed GRB afterglow: non-relativistic case}

\author[D.M. Wei and T. Lu]
{D.M. Wei$^{1,2}$ and T. Lu$^{3,4}$\\
$^{1}$ Purple Mountain Observatory, Chinese Academy of Sciences, Nanjing, 210008, China\\
$^{2}$ National Astronomical Observatories, Chinese Academy of Sciences, China\\
$^{3}$ Department of Astronomy, Nanjing University, Nanjing, 210093, China\\
$^{4}$ LCRHEA, IHEP, CAS, Beijing, China\\}

\maketitle

\noindent{All correspondence please send to:}
\vspace{5mm}

\noindent{D.M. Wei}\\
Purple Mountain Observatory\\
Chinese Academy of Sciences\\
Nanjing, 210008\\
P.R. China

\vspace{5mm}
\noindent{email: dmwei@pmo.ac.cn}\\
fax: 8625-3307381

\newpage

\begin{abstract}

There has been increasing evidence that at least some GRBs are emission beamed. The beamed
GRB afterglow evolution has been discussed by several authors in the ultra-relativistic case.
 It has been shown that the dynamics of the blast wave will be significantly modified by the sideways 
expansion, and there may be a sharp break in the afterglow light curves under certain circumstances. 
However, this is true only when the fireball is still relativistic. Here we present an analytical approach 
to the evolution of the beamed GRB blast wave expanding in the surrounding medium (density $n\propto 
r^{-s}$) in the non-relativistic case, our purpose is to explore whether
the sideways expansion will strongly affect the blast wave evolution as in the relativistic case. 
We find that the blast wave evolution is strongly dependent on the speed of the sideways expansion. 
If it expands with the sound speed, then the jet angle $\theta $ increases with time as $\theta \propto ln t$,
which means that the sideways expansion has little effect on the afterglow light curves, the flux
$F\propto t^{-\frac{3(5\alpha -1)}{5}}$ for $s=0$ and $F\propto t^{-\frac{7\alpha +1}{3}}$ for $s=2$.
It is clear that the light curve of $s=2$ is not always steeper than that of $s=0$, as in the relativistic case.
We also show that if the expansion speed is a constant, then the jet angle $\theta \propto t$, and the radius 
$r \propto t^{0}$, in this case the sideways expansion has the most significant effect on the blast wave evolution, 
the flux $F\propto t^{-(5\alpha -1)}$ independent of $s$, and we expect that there should be a 
smooth and gradual break in the light curve.
\end{abstract}

\begin{keywords}
gamma-rays: bursts
\end{keywords}

\newpage

\section{Introduction}

The fireball model of $\gamma $-ray bursts led to the prediction of the afterglow emission that might be
expected when the energetic shock wave encountered the surrounding medium. The subsequent X-ray
and optical observations of the afterglow appeared to confirm the prediction of the simplest afterglow
model (Wijers, Rees \& Meszaros 1997). This model involved synchrotron emission from electrons
accelerated to a power law energy spectrum in a relativistic blast wave. However, both geometry and
environment can affect the evolution of $\gamma $-ray bursts afterglows (Panaitescu, Meszaros \&
Rees 1998). There has been increasing evidence that at least some GRBs are emission beamed or
the surrounding medium is not uniform.  A class of GRBs whose afterglow exhibited steeper than 
the normal power law decays ($f_{\nu }\propto t^{-2}$) can well be explained by the jet -like geometry 
of the relativistic shock (Rhoads 1997,1999; Sari, Piran \& Halpern 1999; Wei \& Lu 1999), or the
inhomogeneous surrounding medium models (especially the wind model, Dai \& Lu 1998; Chevalier \& Li 1999;
Li \& Chevalier 1999). The jet model is also supported by the steepening of the optical and radio 
light curves seen in GRB 990510 (Stanek et al. 1999; Harrison et al. 1999).

The dynamical evolution of GRB fireballs and the emission features have been studied by many
authors (e.g. Sari 1997; Meszaros, Rees \& Wijers 1998; Wei \& Lu 1998a, 1998b; Sari, Piran \&
Narayan 1998), but most of them considered the fireball being isotropic.
The evolution of the beamed blast wave has been first discussed by Rhoads (1997, 1999). 
He has shown that the lateral expansion of the relativistic plasma causes that at some moment 
the surface of the blast wave starts to increase faster than due to the cone-outflow alone, then
the blast wave begins to decelerate faster than without the sideways expansion since more intersteller
medium has been swept up by the blast wave. He claimed that this effect will produce a sharp break in
the GRB afterglow light curves. More detailed calculation, both numerical and analytical studies
(Moderski, Sikora \& Bulik 1999; Wei \& Lu 1999) have found that unless the opening angle is
very small or that lateral expansion is unimportant, a smooth and gradual transition is expected.
However, these results are valid only when the blast wave is relativistic. As shown by Rhoads (1999)
and Panaitescu \& Meszaros (1999), the Lorentz factor $\Gamma _{b}$ at the radius $r_{b}$ where
the sideways expansion becomes important is $\Gamma _{b}\sim \frac{2}{5\sqrt{3}}\theta _{0}^{-1}$,
where $\theta _{0}$ is the initial half opening angle of the jet. To keep $\Gamma _{b}\gg 1$ requires
$\theta _{0}<0.1$, so if the jet angle is not too small, the blast wave has become non-relativistic when
sideways expansion is important, and the previous results are invalid.

Frail et al. (1999) reported on the results of an extensive observation of the radio afterglow of GRB970508,
lasting 450 days after the burst. They have shown that the spectral and temporal radio behavior 
indicate that the fireball has undergone a transition to non-relativistic expansion at $t\sim 100$ days, 
and they find that the fireball may be initially a wide angle jet of opening angle $\sim 30^{\circ}$.
Therefore it is very interesting and important to study the behavior of the beamed blast wave in the 
non-relativistic case. Recently Huang et al.(1999) have calculated the evolution of jetted GRB ejecta
numerically.

Here we present an analytical approach to the evolution of the beamed GRB afterglow
in the non-relativistic case, including both homogeneous medium and wind-shaped medium
(the density $n\propto r^{-2}$), the main purpose is to explore whether the sideways expansion 
will strongly modify the blast wave evolution and the afterglow light curve behavior as in the
relativistic case. In next section we consider the dynamical evolution of blast wave, in section 3
we calculate the jet emission and afterglow light curve analytically, and finally we give some
discussions and conclusions.

\section{Dynamical evolution of the jet}

Now we consider the evolution of an adiabatic blast wave expanding in surrounding medium. Assuming 
that the medium density $n\propto r^{-s}$, $s=0$ corresponds to the homogeneous medium, and $s=2$
corresponds to the wind-shaped medium. For completeness, here we first outline the results of evolution
in the relativistic case.

\subsection{relativistic case}

For energy conservation, the evolution equation of blast wave is
\be
\Gamma ^{2}N=const
\ee
where $\Gamma $ is the bulk Lorentz factor, and $N$ is the total baryon numbers swept up by the blast
wave, $N\propto r^{3-s}(1-cos\theta _{j})\propto r^{3-s}\theta _{j}^{2}$ for $\theta _{j}\ll 1$, and $\theta _{j}
=\theta _{0}+\theta '$,  where $\theta _{0}$ is the initial jet opening half-angle, $\theta '$ describes the 
lateral expansion, which can be simply written as $\theta '\sim v_{s}t_{co}/ct$, $v_{s}$ is the expanding 
velocity of ejecta material in its comoving frame, and $t\,(t_{co})$ is the time measured in the burster
frame (comoving frame). For relativistic expanding material it is appropriate to take $v_{s}$ to be the 
sound speed $v_{s}=c/\sqrt{3}$. Since the jet expands relativistically, there is the relation $T\propto 
r/\Gamma ^{2}$, where $T$ is the time measured in the observer frame. According to Wei \& Lu (1999),
for $s=0$, we obtain
\be
\Gamma \propto \left \{
      \begin{array}{cc}
          T^{-3/8}[1+(\frac{T}{T_{b}})^{3/8}]^{-1/4}, & {\rm if}\,\, T<T_{b}\\
          T^{-3/8}[1+(\frac{T}{T_{b}})^{1/2}]^{-1/4}, & {\rm if}\,\, T >T_{b}
      \end{array}
  \right .
\ee
where $T_{b}$ is the time at that moment the sideways expansion is important. Similarly, for $s=2$,
we have
\be
\Gamma \propto \left \{
      \begin{array}{cc}
          T^{-1/4}[1+(\frac{T}{T_{b}})^{1/4}]^{-1/2}, & {\rm if}\,\, T<T_{b}\\
          T^{-1/4}[1+(\frac{T}{T_{b}})^{1/2}]^{-1/2}, & {\rm if}\,\, T >T_{b}
      \end{array}
  \right .
\ee
so we see that, for $T\ll T_{b}$, $\Gamma \propto T^{-3/8}$ and $r\propto T^{1/4}$ for $s=0$, and 
$\Gamma \propto T^{-1/4}$ and $r\propto T^{1/2}$ for $s=2$, while for $T\gg T_{b}$, both $s=0$ and 
$s=2$ give $\Gamma \propto T^{-1/2}$ and $r\propto T^{0}$.

\subsection{non-relativistic case}

The evolution of the jetted blast wave in the non-relativistic case is
\be
\beta ^{2}N=const
\ee
where $\beta $ is the velocity of the blast wave in units of light speed $c$. Thus, for $\theta _{j}\ll 1$, 
we have $\beta ^{2}r^{3-s}\theta _{j}^{2}=const$. In this case the jet angle $\theta_{j} $
can be written as $\theta_{j} =\theta _{0}+\int \frac{\beta _{s}dt}{\int \beta dt}$, where $\beta _{s}$ is the spread 
velocity of ejecta material in units of light speed $c$, which cannot be simply determined. Here we 
consider several situations.

(i) {\em $\beta _{s}$=sound speed} {\hspace{3mm}} Kirk \& Duffy (1999) have derived the sound speed in the fluid
\be
c_{s}^{2}=\frac{\hat {\gamma }P}{\rho }[\frac{(\hat {\gamma }-1)\rho }{(\hat {\gamma }-1)\rho+\hat {\gamma }P}]
\ee
where $P$ is the pressure, $\rho $ is the mass density, and $\hat {\gamma }$ is the adiabatic index.
According to Huang et al. (1999), the sound speed can be written as
\be
c_{s}^{2}=\frac{\hat {\gamma }(\hat {\gamma }-1)(\gamma -1)}{1+\hat {\gamma }(\gamma -1)}
\ee
In the non-relativistic limit ($\gamma \simeq 1,\,\hat {\gamma }\simeq 5/3$), one gets $c_{s}=\sqrt{5}\beta /3$,
then we obtain the jet angle $\theta_{j} =\theta _{0}+\frac{2\sqrt{5}}{3(5-s)}\ln(\frac{t}{t_{0}})$, where $t_{0}\approx 
t_{NR}$, $t_{NR}$ is the time when the blast wave turns from relativistic to non-relativistic. Therefore we 
see that since the jet angle $\theta _{j}$ increases with time as the $\log$ relation, it will not strongly 
affect the evolution of the jet, the variation of the jet velocity is about $\beta \propto t^{-(3-s)/(5-s)}$,
i.e., $\beta \propto t^{-3/5}$ for $s=0$, and $\beta \propto t^{-1/3}$ for s=2.

(ii) {\rm $\beta _{s}=const$  {\hspace{3mm}} If the jet spreads at a constant speed, then for $t\gg t_{NR}$, 
the angle $\theta _{j}\simeq \theta _{0}+\frac{2}{3-s}\frac{\beta _{s}}{\beta }$, and the evolution of $\beta $ satisfies
$\beta ^{5-s}(\theta _{0}+\frac{\beta _{s}}{\beta })^{2}\propto t^{-(3-s)}$, so when $\beta \ll \frac{2}{3-s}\beta _{s}\theta _{0}^{-1}$,
we get $\beta \propto t^{-1},\,r\sim const$ independent of $s$. We see that in this case it is similar to
the situation that sideways expansion is important in the relativistic case.

(iii) {\rm In general}, we assume that the ratio of the jet spread velocity and the blast wave velocity
varies with time, $\beta _{s}/\beta =at^{q }$, then we obtain when $t \gg t_{NR}$, the jet angle
$\theta _{j}\simeq \theta _{0}+\frac{2a}{q(5-s) }t^{q}$, and for $t \gg ((5-s)q \theta _{0}/2a)^{1/q }$,
the evolution of $\beta $ satisfies $\beta \propto t^{-\frac{3-s+2q }{5-s}}$, so $\beta \propto t^{-
\frac{3+2q }{5}}$ for $s=0$ and $\beta \propto t^{-\frac{1+2q }{3}}$ for $s=2$. We see that 
$q =1$ corresponds to the case (ii). Therefore it seems that the evolution of the blast wave
strongly depend on the parameter $q $.

\section{The emission from the jet}

Now we calculate the emission flux from the non-relativistic jet. Here we adopt the formulation and
notations of Mao \& Yi (1994). In our model the ejecta is flowing outwards in a cone with opening
half angle $\theta _{j}$. For simplicity, we assume that the radiation is isotropic in the comoving
frame of the ejecta and has no dependence on the angular positions within the cone. The radiation 
cone is uniquely defined by the angular spherical coordinates ($\theta ,\,\phi $), here $\theta $ is
the angle between the line of sight (along $z$-axis) and the symmetry axis, and $\phi $ is the azimuthal
angle. Because of cylindrical symmetry, we can assume that the symmetry axis of the cone is in the 
$y-z$ plane. In order to see more clearly, let us establish an auxiliary coordinate system ($x', y', z'$) 
with the $z'$-axis along the symmetry axis of the cone and the
$x'$ parallel the $x$-axis. Then the position within the cone is specified by its angular spherical 
coordinates $\theta '$ and $\phi '$ ($0\leq \theta '\leq \theta _{j}$, $0\leq \phi '\leq 2\pi $). It can be shown that
the angle $\Theta $ between a direction ($\theta ',\,\phi '$) within the cone, and the line of sight satisfies
$cos\Theta =cos\theta cos\theta '-sin\theta sin\theta 'sin\phi '$. Then the observed flux is
\be
F(\nu ,\theta )=\int_{0}^{2\pi }d\phi '\int_{0}^{\theta _{j}}sin\theta 'd\theta ' cos\Theta D ^{3}
I'(\nu D^{-1})\frac{r^{2}}{d^{2}}
\ee
where $D=[\Gamma (1-\beta cos\Theta )]^{-1}$ is the Doppler factor,  $\beta =(1-\Gamma ^{-2})^{1/2}$, 
 $\nu =D\nu '$, $I'(\nu ')$ is the specific intensity of synchrotron radiation at $\nu '$, and $d$ is the 
distance of the burst source. Here the quantities with prime are measured in the comoving frame. For
simplicity we have ignored the relative time delay of radiation from different parts of the cone.

For the non-relativistic blast wave, it is widely believed that the accelerated electrons have power-law 
energy distribution $n(\gamma )\propto \gamma ^{-p}$ in the range $\gamma _{1}\leq \gamma \leq 
\gamma _{2}$, then the typical electron energy $\gamma _{1}\simeq \frac{1}{2}\epsilon _{e}\frac{p-2}
{p-1}\frac{m_{p}}{m_{e}}\beta ^{2}$, where $\epsilon _{e}$ is the energy fraction occupied by the 
electrons, $m_{p}$ and $m_{e}$ are the mass of proton and electron. Assuming that $\epsilon _{e}$ 
is a constant, then we have $\gamma _{1}\propto \beta ^{2}$. For the non-relativistic case, $\Gamma \sim 1$,
then $r=D\Gamma \beta ct\propto \beta t/(1-\beta cos\Theta) ,\,r'=D\beta ct\propto \beta t/(1-\beta cos\Theta) $,
the energy density $u\propto n\beta ^{2}$, the magnetic field strength $B\propto u^{1/2}\propto \beta n^{1/2}$,
the peak frequency of synchrotron radiation $\nu _{m}=D\nu _{m}'\propto D\gamma _{1}^{2}B\propto 
\beta ^{5}n^{1/2}/(1-\beta cos\Theta) $. Assuming that the emission spectrum $I'(\nu ')\propto \nu '^{-\alpha }$, then
$I'(\nu ')=I'(\nu _{m}')(\frac{\nu '}{\nu _{m}'})^{-\alpha }=I'(\nu _{m}')(\frac{\nu }{\nu _{m}})^{-\alpha }
\propto \beta ^{2+5\alpha }n^{(3+\alpha )/2}t\nu ^{-\alpha }/(1-\beta cos\Theta )^{1+\alpha }$. Therefore
we have the emission flux 
\be
F(\nu ,\theta )\propto \nu ^{-\alpha }\beta ^{4+5\alpha }n^{(3+\alpha )/2}t^{3}g(\theta ,\beta ,\alpha )
\ee
where
\be
g(\theta ,\beta ,\alpha )=\int_{0}^{2\pi }d\phi '\int_{0}^{\theta _{j}}sin\theta 'd\theta ' cos\Theta 
(1-\beta cos\Theta )^{-(6+\alpha )}
\ee
In general, the value of $g$ can only be calculated numerically. However here we consider the case
$\theta _{j}\ll 1$ and $\theta \ll 1$, then $cos\Theta \approx cos\theta cos\theta '$. In this case we
can calculate the value of $g$ analytically under certain conditions. After complicated calculation
we find $g\approx \pi cos\theta sin^{2}\theta _{j}$, so finally we get
\be
F(\nu ,\theta )\propto \nu ^{-\alpha }\beta ^{4+5\alpha }n^{(3+\alpha )/2}t^{3}cos\theta sin^{2}\theta _{j}
\ee

From the previous section, we can obtain the GRB afterglow light curves according to different spread 
velocity. For case (i), we find $F(\nu ,\theta )\propto t^{3-\frac{(3-s)(4+5\alpha )+s(3+\alpha )}{5-s}}cos\theta 
sin \theta _{j}^{2}$, more clearly, $F\propto t^{-\frac{3(5\alpha -1)}{5}}cos\theta sin^{2}\theta _{j}$ for $s=0$,
and $F\propto t^{-\frac{(7\alpha +1)}{3}}cos\theta sin^{2}\theta _{j}$ for $s=2$. For case (ii), we find it 
is simple, $F\propto t^{-(5\alpha -1)}cos\theta $, independent of $s$. For case (iii), it is shown that
$F\propto t^{3+2q-\frac{(3-s+2q)(4+5\alpha )+s(1-q)(3+\alpha )}{5-s}}cos\theta $, for $s=0$ $F\propto 
t^{-\frac{(3+2q)(5\alpha -1)}{5}}cos\theta $, and for $s=2$ $F\propto t^{3+2q-\frac{(1+2q)(4+5\alpha )+
2(1-q)(3+\alpha )}{3}}cos\theta $. It is obviously that when $q=0$, it reduces to case (i), and when
$q=1$, it reduces to case (ii).

\section{Discussion and conclusions}

In this paper we have investigated the dynamical evolution of the non-relativistic blast wave in the 
surrounding medium (density $n\propto r^{-s}$), especially whether the sideways expansion will
have great effect on the blast wave evolution and the GRB afterglow light curves. 
We find that whether the sideways expansion being important is strongly dependent on the spread 
velocity of the jetted material. We think that it is more reasonble for material to spread with sound
speed, and in this case, the jet angle $\theta _{j}\propto \ln t$, so it has little effect on the GRB 
afterglow light curves, i.e. the light curves will be nearly a simple power law without a break, but
the increase of $\theta _{j}$ cannot be ignored, the blast wave will approach nearly spherical after
certain time.

In the limiting case where the spread velocity is a constant, the sideways expansion is very important
as in the relativistic case, $\theta _{j}\propto t$ and $r$ is nearly a constant, which is independent of $s$.
In this case there should be a break in the GRB afterglow light curves. However, we note that the above 
results is valid only when $t\gg t_{NR}$, and $t_{NR}$ is usually about one month or even more, so 
we expect that the afterglow light curve should become steeper smoothly and gradually.  As for the more
general situation, we find that the light curve behavior should lie between the above two cases.

Here we calculated the blast wave evolution and the jet emission features in the medium $n\propto r^{-s}$,
and especially we compare the results of $s=0$ and $s=2$. It is well known that in the relativistic case,
the light curve of $s=2$ is steeper than that of $s=0$, however, here we find that in the non-relativistic 
case this is not true. The slope of the light curve also depends on the spectral index $\alpha $. For 
example, in case (i), we find that when $\alpha <1.4$, the light curve of $s=2$ is steeper than that of
$s=0$, while if $\alpha >1.4$, the situation is opposite. But for case (ii), i.e. the sideways expansion
is very important, we find that the light curve is independent of $s$, since in this case $r\sim constant$.

Recently Frail et al. (1999) have reported on the results of 450 days observations of the radio afterglow 
of GRB 970508, and indicated that the fireball has undergone a transition to non-relativistic expansion 
at $t\sim 100$ days. Therefore it is very interesting and important to explore the dynimacal evolution 
of the non-relativistic jetted material in various surrounding medium, since we expect that over the duration 
of the radio emission, the emitting material typically become non-relativistic.

\section{Acknowledgements}
This work is supported by the National Natural Science Foundation (19703003 and 
19773007) and the National Climbing Project on Fundamental Researches of China.


\begin{thebibliography}{}

\bibitem [ ]{} Chevalier, R.A., Li, Z.Y., 1999, ApJ, 520, L29
\bibitem [ ]{} Dai, Z.G., Lu, T., 1998, MNRAS, 298, 87
\bibitem [ ]{} Frail, D.A., Waxman, E., Kulkarni, S.R., 1999, astro-ph/9910319
\bibitem [ ]{} Harrison, F.A., etal., 1999, ApJ, 523, L121
\bibitem [ ]{} Huang, Y.F., Gou, L.J., Dai, Z.G., Lu, T., 1999, astro-ph/9910493
\bibitem [ ]{} Kirk, J.G., Duffy, P., 1999, astro-ph/9905069
\bibitem [ ]{} Li, Z.Y., Chevalier, R.A., 1999, astro-ph/9903483
\bibitem [ ]{} Mao, S., Yi, I., 1994, ApJ, 424, L131
\bibitem [ ]{} M\'{e}sz\'{a}ros, P., Rees, M.J., Wijers, R., 1998, ApJ, 499, 301
\bibitem [ ]{} Moderski, R., Sikora, M., Bulik, T., 1999, astro-ph/9904310
\bibitem [ ]{} Panaitescu, A., M\'{e}sz\'{a}ros, P., Rees, M.J., 1998, ApJ, 503, 314
\bibitem [ ]{} Panaitescu, A., M\'{e}sz\'{a}ros, P., 1999, astro-ph/9806016
\bibitem [ ]{} Rhoads, J.E., 1997, ApJ, 487, L1
\bibitem [ ]{} Rhoads, J.E., 1999, astro-ph/9903399
\bibitem [ ]{} Sari, R., 1997, ApJ, 494, L49
\bibitem [ ]{} Sari, R., Piran, T., Narayan, 1998, ApJ, 497, L17
\bibitem [ ]{} Sari, R., Piran, T., Halpern, J.P., 1999, ApJ, 519, L17
\bibitem [ ]{} Stanek, K.Z., et al., 1999, ApJ, 522, L39
\bibitem [ ]{} Wei, D.M., Lu, T., 1998a, ApJ, 499, 754
\bibitem [ ]{} Wei, D.M., Lu, T., 1998b, ApJ, 505, 252
\bibitem [ ]{} Wei, D.M., Lu, T., 1999,  ApJ, in press (astro-ph/9908273)
\bibitem [ ]{} Wijers, R.A.M.J., Rees, M.J., M\'{e}sz\'{a}ros, P., 1997, MNRAS, 288, L51

\end{thebibliography}
\end{document}